\documentclass[conference]{IEEEtran}
\IEEEoverridecommandlockouts
\usepackage{cite}
\usepackage{amsmath,amssymb,amsfonts}
\usepackage{algorithmic}
\usepackage{graphicx}
\usepackage{textcomp}
\usepackage{xcolor}
\usepackage{subfig}
\begin{document}

\title{Evaluation of RF Wireless Power Transfer for Low-Power Aircraft Sensors\\}
\author{\IEEEauthorblockN{Jan Tepper\IEEEauthorrefmark{1},
Aygün Baltaci\IEEEauthorrefmark{1}, Bernd Schleicher\IEEEauthorrefmark{1}\IEEEauthorrefmark{2}, Andreas Drexler\IEEEauthorrefmark{1}\IEEEauthorrefmark{4}, Svetoslav Duhovnikov\IEEEauthorrefmark{1}, Mustafa Ozger\IEEEauthorrefmark{3},\\ Morteza Tavana\IEEEauthorrefmark{3}, Cicek Cavdar\IEEEauthorrefmark{3} and
Dominic Schupke\IEEEauthorrefmark{1}}
\IEEEauthorblockA{\IEEEauthorrefmark{1}  Airbus, Central Research and Technology, Communication Technologies, Ottobrunn/Munich, Germany \\
\IEEEauthorrefmark{2} Now with Qorvo Munich GmbH, Munich, Germany
 \\
 \IEEEauthorrefmark{4} Now with Schalt-Technik-Huber GmbH, Munich, Germany
 \\
\IEEEauthorrefmark{3} School of Electrical Engineering and Computer Science, KTH Royal Institute of Technology, Stockholm, Sweden \\
Email: \{jan.tepper, dominic.schupke\}@airbus.com}}

\maketitle

\begin{abstract}
Low-power sensors can be integrated into an aircraft for numerous use cases. Conventionally, these sensors are powered via cables, which introduces various disadvantages to the overall efficiency of the aircraft. Alternatively, batteries may be used. However, this implies the necessity of additional maintenance for battery replacement. Another option to remove power cables is to use Radio Frequency (RF) Wireless Power Transfer (WPT) systems. Although general RF WPT technology has been studied in the literature, its feasibility for aviation use cases is not fully investigated. In this paper, we study the feasibility of RF WPT to wirelessly power low-power in-cabin sensors. In a cabin mock-up we show that RF WPT techonology is capable of almost fully covering an area of 20 seats and quantitatively assess this using Received Signal Strength Indicators (up to $28$\,mW) and packet interval rate (up to $5.5$\,Hz). Furthermore, we perform multi-tone sinusoidal wave experiments for power transmission scheme in a lab environment and thereby present potential ways to improve receiver sensitivity and consequently increase the WPT coverage in the cabin without changing the average transmission power. The overall results show that certain low-power cabin use cases can be supported by already existing commercial RF WPT systems. 
\end{abstract}


\section{Introduction}
Low-power sensors have become an integral part of modern commercial aircrafts and can aid the aircraft operations in a plethora of use cases. Such instances can be cabin temperature monitoring, seat applications, brake-condition monitoring, structural health and many more. The number of sensors can be as high as $4,000$ in a large aircraft \cite{b8}. However, powering such a large number of sensors via cables leads to a number of disadvantages such as increased weight, fuel consumption and CO$_2$ emission. As of today, the total weight of wires can be as high as $5,700$ kg on an Airbus A380-800 \cite{b9}. Additionally, cables and harnesses account for significant efforts in the final assembly line. Furthermore, cabin modifications such as seat rearrangements become more complex if rewiring needs to be considered. Therefore, wireless technologies are recently favored to overcome the aforementioned issues in the next-generation aircraft. 

Cable reduction is already addressed for communications, for instance, by IoT type applications or,  related to safety and regulatory of flight, by Wireless Avionics Intra Communications (WAIC) applications. The usage of wireless technologies can be further extended to remote powering scenarios to reduce the number of powering cables. One method can be to utilize RF-based WPT systems. 

Designing an active WPT scheme onboard an aircraft can not only allow for the reduction of cables, but also target the powering of wireless sensors at certain positions that are difficult to access using cables, such as the landing gear sensors. Therefore, a number of aircraft use cases already exist to evaluate the performance of WPT systems. On the other hand, the main drawbacks of WPT are poorer power efficiency as well as more strict regulatory constraints with respect to the allowed RF transmit power and operation frequency compared with conventional electrical cabling. 

In this paper, we experimentally assess the feasibility of WPT for low-power cabin sensors onboard an aircraft. A WPT system is set up inside an Airbus A330 cabin mock-up. The achievable Received Signal Strength Indicator (RSSI) and resulting WPT coverage are measured along different locations in the cabin.
Moreover, we investigate the RF to DC conversion efficiency gain on the WPT RX side by increasing the number of sinusoidal tones in the RF waveform in a laboratory environment. We estimate the resulting improvements in RX sensitivity, i.e. the capability to operate at low power, which would allow to reduce transmission power or to increase the WPT coverage inside an aircraft cabin. 

The rest of the paper is organized as follows. In Section II, recent works in the scope of RF WPT are provided. Afterwards, we discuss the regulatory aspects of realizing WPT systems for the next-generation aircraft in Section III. Section IV describes our measurement setup in detail. The measurement results are presented in Section V, and the conclusions regarding the feasibility of WPT systems for aircraft use cases are drawn in Section VI.  

\section{Previous work}
In general, WPT can be subdivided into radiative wireless charging and non-radiative wireless charging. The RF approach presented in this paper falls in the category of radiative charging and is more suitable for a cabin environment in which distances of several meters need to be overcome.
While the first microwave powered system was demonstrated in 1964 by Brown \cite{Brown}, RF WPT became more relevant with the rise of portable electronic devices in the 90s and the advent of IoT and low-power sensors nowadays. Commercial options are nowadays available for instance through the RF-based COTA technology from OSSIA \cite{OSSIA} or buying RF WPT hardware from Powercast\cite{Powercast} including wireless transmitters and receivers.
The topic is part of current research, investigating ways to improve efficiencies, exploring new use cases, or the integration into communication systems.  Recently, La Rosa et al \cite{LaRosa} demonstrated a silicon-based integrated circuit for RF energy harvesting with an improved sensitivity down to -18.8\,dBm input power at 900\,MHz. Such a low sensitivity enables the charging of capacitors or batteries in the presence of very weak RF fields. WPT has also been considered for the powering of Body Area Network (BAN) devices for biomedical applications \cite{Xia2014}. In this work, the wireless charging has been further refined through a tracking loop, which adjusts the RF to DC rectifiers threshold to achieve maximum efficiency. Additionally, the coexistence of wireless communications has been studied in terms of minimizing interference \cite{Shinohara2014} as well as using a single link to perform the two functions \cite{Mao2017}. A comprehensive overview of the history and state of the art of remote powering is given in \cite{Lu2016}.
%To our knowledge, the topic of RF WPT has not been applied in the context of avionic cabin applications, which will be the scope of this work.

\section{Regulations of Wireless Power Transfer}
WPT systems are less considered in aircraft use cases so far. In this aspect, it is significant to evaluate the regulatory aspects to ensure the compliance of such a system for an aircraft. The regulations can be essential, especially in terms of the the transmit power levels as well as the operation frequency. Therefore, we will briefly provide the current regulations in the U.S. and Europe. The reader may refer to the cited documents for further guidance.

The European authority for frequency regulation is the ECC. According to the ECC regulations \cite{b2}, the bands which allow relatively high powers are $865-868$\,MHz ($2$\,W ERP) and $915-921$\,MHz ($4$\,W ERP). For these bands, an additional ETSI (European Telecommunications Standards Institute) standard also needs to be taken into account \cite{b3}. The U.S. counterpart of the ECC is the FCC. In FCC regulations \cite{b1}, the highest output power in the bands $902-928$\,MHz and $2400-2483.5$\,MHz is $4$\,W including an antenna gain of $6$\,dBi.

Another essential aspect is the RF exposure to passengers and the crew inside the aircraft. The International Commission on Non-ionizing Radiation Protection (ICNIRP) Guidelines regulates these aspects \cite{b4}. A maximum power density limit of $f/200$\,W/m$^2$ ($f$ in MHz) for frequencies $f$ of $0.4 - 2$\,GHz and $10$\,W/m$^2$ for frequencies $>2$\, GHz is stated. 
Overall, unharmonized nation-wide regulations pose difficulties for the WPT integration on aircrafts on international routes and harmonization efforts may be required to fully unfold its potential.

\section{Measurement setup}
In this paper, two aspects of WPT systems are experimentally investigated. First, the achievable received power levels and WPT coverage are assessed. Second, the improvement of RF to DC power conversion is addressed through the use of multi-tone sinusoidal RF transmission.

\subsection{Cabin Mock-up Measurements}  

The measurements for achievable power level and coverage are conducted in a mock-up A330 cabin. As for the WPT system, a commercial Powercast 2110 evaluation kit is used \cite{b10}. It consists of a power TX board, a power RX board, a ZigBee sensor as data TX and another ZigBee sensor as data RX. A development board is used to attach the ZigBee RX to a PC to read the received data, which includes charging time, RSSI and other sensor readings. The hardware setup is shown in Figure 1. The sender operates at $915$ MHz with $3$\,W EIRP of transmission power. The signal modulation is DSSS and the TX has an omni-directional antenna with a beam divergence of 60° in both horizontal and vertical direction. 

The RX board converts the RF signal to DC. It then charges a capacitor until a certain threshold. Afterwards, the output voltage is boosted to 3.3\,V, which can then be provided to the ZigBee sensor. The RF to DC conversion efficiency is an essential parameter in the overall system and is studied in the second part of the paper. The ZigBee sensor is the WSN-EVA-01 board, which reads the RSSI, packet interval and other sensor information such as temperature and humidity. RSSI is measured before RF to DC conversion and is given mW. Packet interval and RSSI are used to assess the performance of the WPT system. When the sensor then draws current, the voltage drops and the recharging repeats over. The time it takes for one such cycle is referred as the packet interval. Finally, The ZigBee RX reads the packet sent by the TX and reports them to the PC.

To further illustrate, Figure \ref{scope} shows the loading curve of the capacitor when the board is fed with a sinusoidal signal at $915$ MHz with an input power of $2$ dBm. Two regions can be identified in the loading curve. In the power-up region, the capacitor is first completely discharged. As the board is fed with an input signal, the voltage across the storage capacitor increases until it reaches a certain threshold at about $1.27$\,V. Once this threshold is reached, the output voltage is boosted to $3.3$\,V by another DC-DC converter and then supplied to the ZigBee sensor. 

In the operating region, the voltage drops as the sensor draws current, wakes up, sends packets and sleeps. After a slight dip, the capacitor is recharged until the voltage threshold, the sensor sends packets again and this process is repeated continuously. The frequency of this cycle depends on the amount of received RF power and the efficiency of the RF to DC conversion. This frequency is referred as the packet interval and its inverse is the operation interval.

\begin{figure}[htbp]
\centerline{\includegraphics[width=0.5\textwidth]{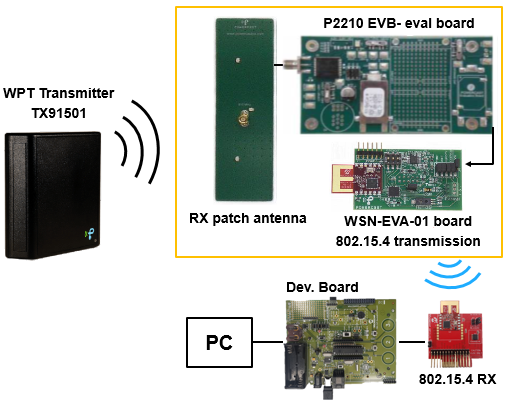}}\label{setup1}
\caption{Sketch of the setup for testing the WPT. WPT is indicated in black, wireless data transmission is indicated in blue.}
\label{fig}
\end{figure}

In the mock-up cabin, the WPT TX is placed at two locations, one at the cabin ceiling with the antenna boresight facing downwards (referred to as Loc.\,1) and once at the cabin wall with the antenna boresight facing perpendicular away from the wall (referred to as Loc.\,2). Pictures of the setup are shown in Fig.\,\ref{mockup}. The TXs are placed at such different orientations to observe its effect on the WPT coverage in the cabin. 
\begin{figure}[htbp]
\centerline{\includegraphics[width=0.5\textwidth]{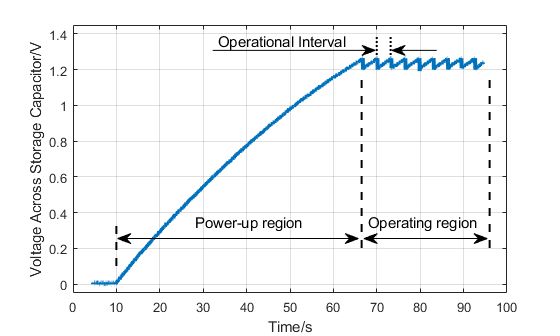}}
\caption{Charging curve of the capacitor on the P2210 Board after RF-DC conversion.}
\label{scope}
\end{figure}

\begin{figure}[htbp]
\centerline{\includegraphics[width=0.5\textwidth]{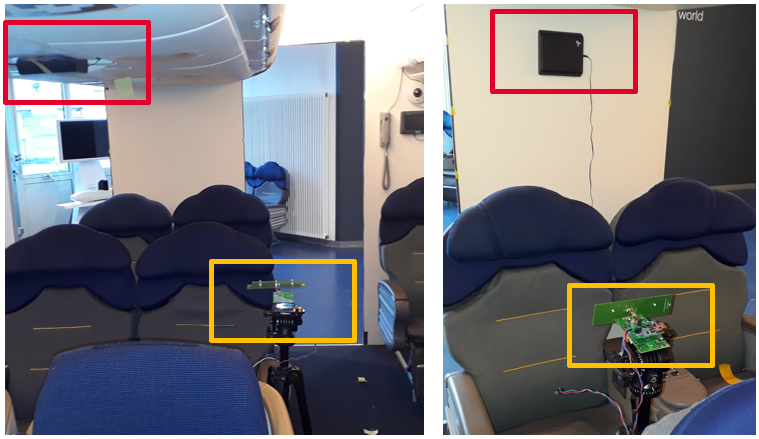}}\label{setup1}
\caption{Pictures of the two TX locations marked in red: Loc. 1 on the left and Loc. 2 on the right. Additionally, two exemplary RX locations are shown in orange. }
\label{mockup}
\end{figure}

\begin{figure}[htbp]
\centerline{\includegraphics[width=0.4\textwidth]{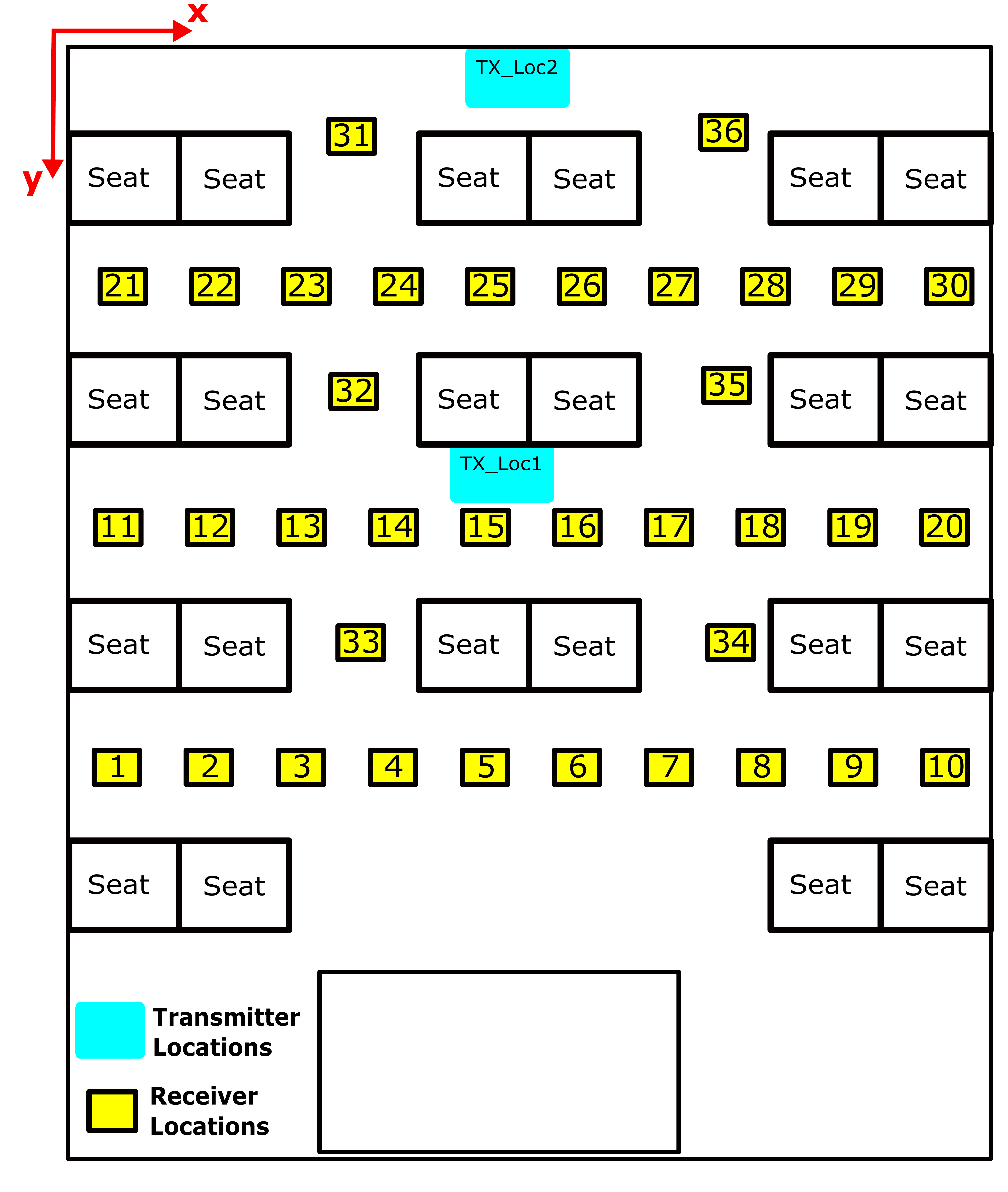}}
\caption{Top view of the cabin mock-up showing the 36 tested RX locations and the two TX locations and indicating the coordinate system.}
\label{rx_pos}
\end{figure}

The power transfer is measured at $36$ different RX points at each TX location. The RX positions are shown in the top view of the cabin in Fig.\,\ref{rx_pos}. These location numbers are also used in Fig. \ref{appendix-table}.

\subsection{Laboratory Measurements}
\begin{figure}[htbp]
\centerline{\includegraphics[width=0.5\textwidth]{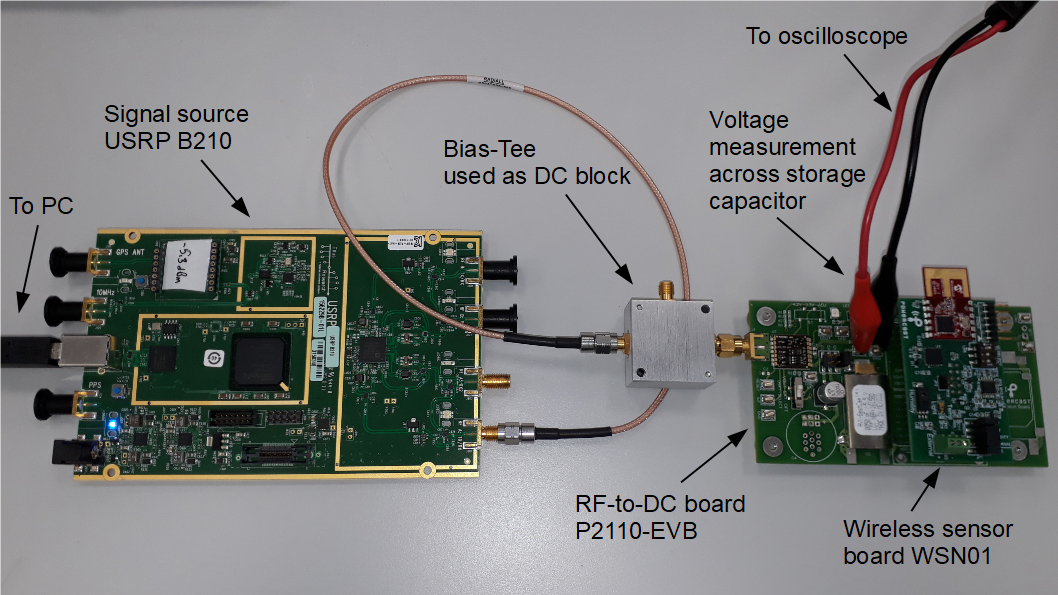}}
\caption{Picture of the multi-tone setup.}
\label{setup-multitone}
\end{figure}

These measurements are conducted to test how the number of sine-wave tone on the TX signal contributes to the RX sensitivity for a given average TX power. The setup is shown in Fig.\,\ref{setup-multitone}. An SDR board, ETTUS USRP B210\cite{b5}, is used as an RF power TX and the output sinusoidal signal is generated using GNU radio\cite{b6} software on a PC. The number of tones on the generated sine-wave is varied between $1-5$ to observe its effect on the RX sensitivity. The power RX board is connected to the SDR via an RF cable to avoid external effects during the measurements. The charging of the RX board is monitored by tapping the signal at its capacitor and displaying the voltage on an oscilloscope. 

Again, we use the aforementioned RSSI and packet interval frequency as an easily accessible measure for the conversion efficiency when the average input power is held constant. The SDR provides the option to create varying-tone sinusoidal waveforms with equal average power so that the effect of the conversion efficiency can be studied.

\section{Results}

In this section, the results of the mock-up cabin measurements as well as the laboratory measurements are presented. 

\subsection{Cabin Mockup Tests}
\begin{figure}[!ht]
   \centering
   \subfloat{\includegraphics[width=.4\textwidth]{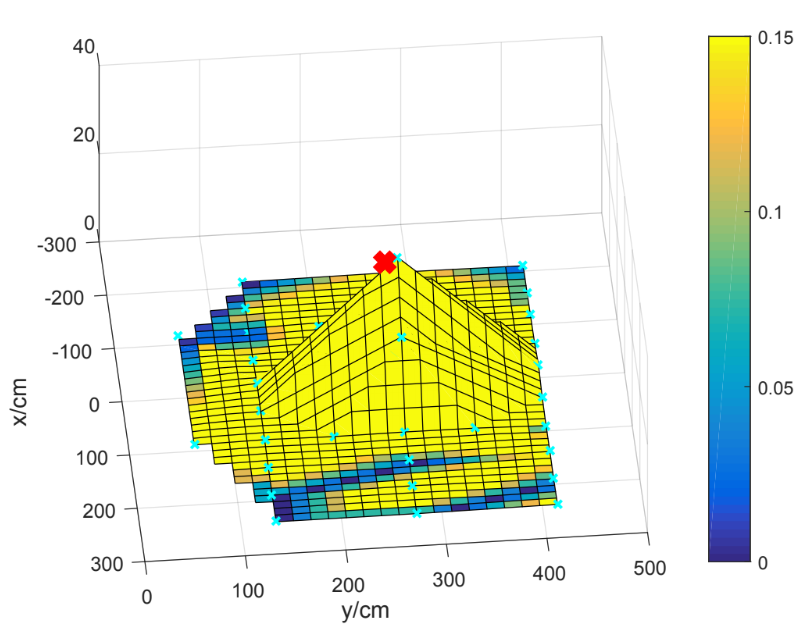}}
  \\
 \quad \subfloat{\includegraphics[width=.4\textwidth]{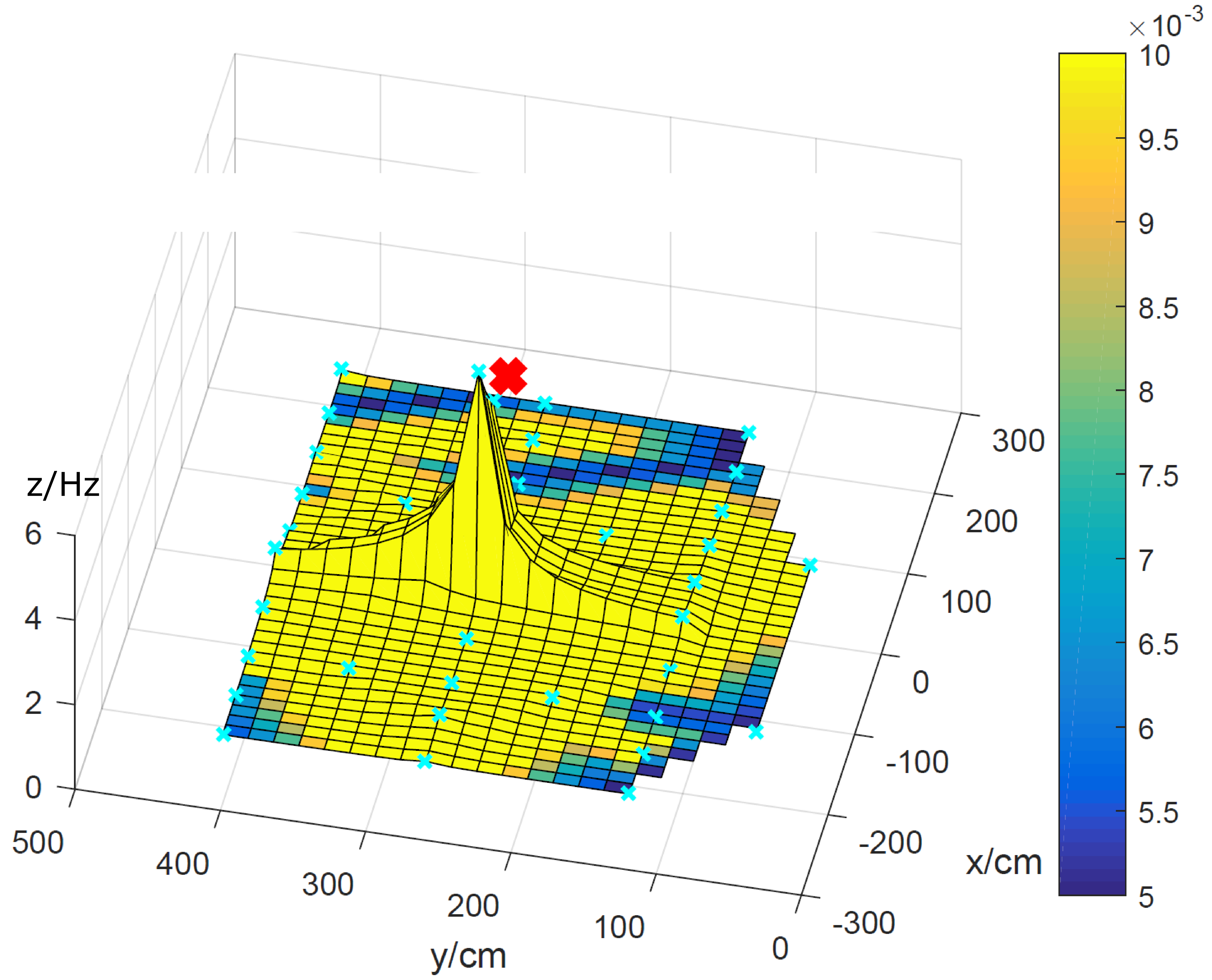}}
   \caption{Received power levels in mW (top) and packet rate frequency in Hz (bottom) for TX location 1. The red cross indicates the TX position. Note that the point of view is not the same for the plots. Also, the colorbar is adjusted such that areas without reception are highlighted in blue. X and Y axis are interpolated by a factor of 4.}
   \label{TX1-results}
\end{figure}

\begin{figure}[htbp]
\centerline{\includegraphics[width=0.5\textwidth]{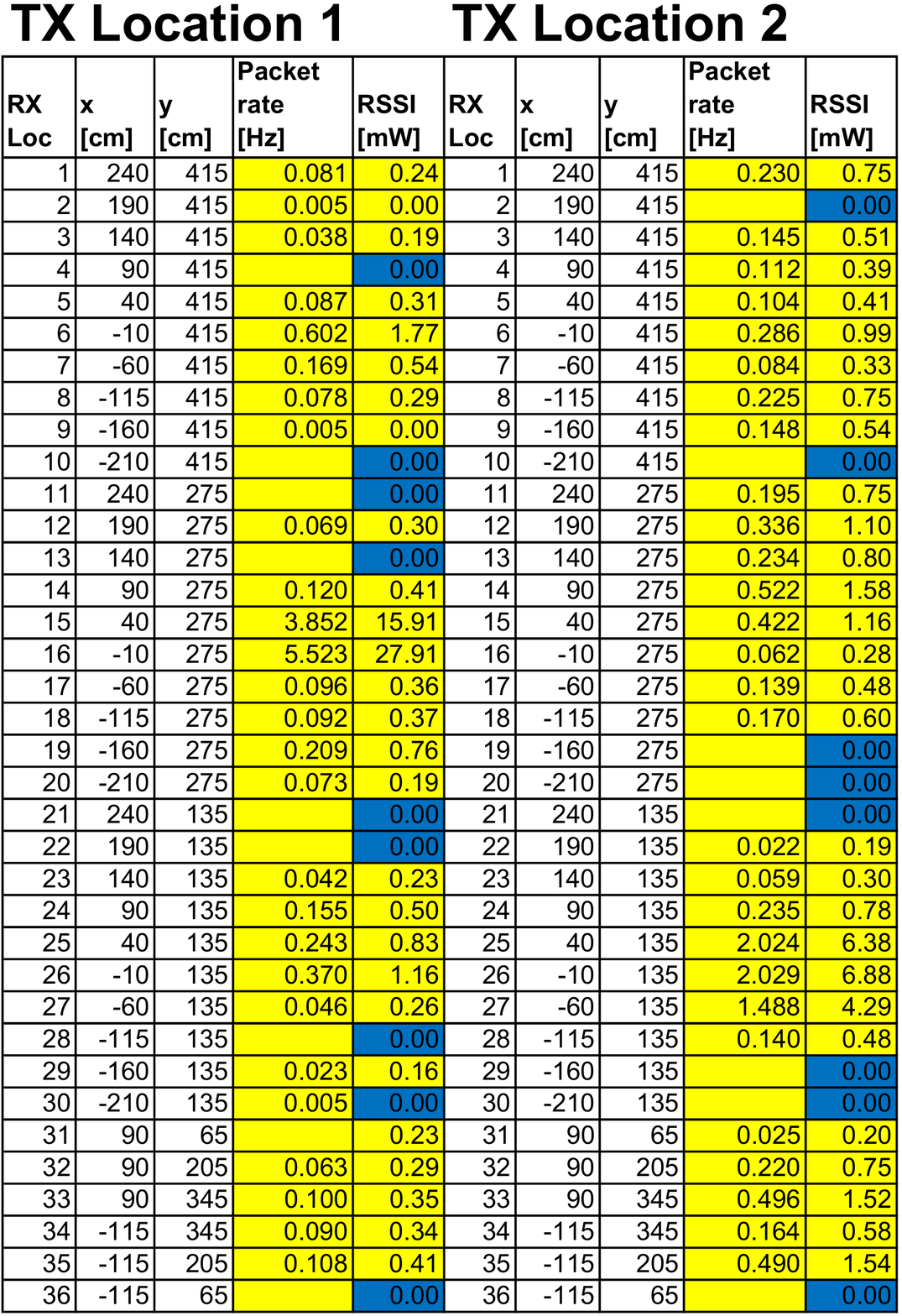}}\label{table-appendix}
\caption{All measured RSSI and packet frame rates and the corresponding coordinates of the RX for the two TX locations. Cells marked in blue did not receive any power.}
\label{appendix-table}
\end{figure}

The results for the received power levels and packet frame rate are shown in Fig.\,\ref{TX1-results}. The red cross shows the TX location and the light turquoise ones show the RX locations. Data interpolation is applied between the RX locations to generate the graph. The exact coordinates of each measurement point, their separation distance as well as the corresponding numerical measurement results can be found in Fig. \ref{appendix-table}.

For TX location 1, the RX module receives sufficient power to overcome the operation threshold for the low-power ZigBee in $27$ RX locations out of $36$. The RSSI drops around the corners of the mock-up. It is significant to observe the unsymmetrical distribution of the RSSI in the graphs. This can be due to the strong fading and multipath effects in the cabin. The maximum RSSI is measured to be around $27.91$ mW at RX location $16$, which is almost right underneath the TX. The packet interval rate also has a similar behavior but its curve (Fig. 6 bottom graph) is steeper compared with the RSSI (Fig. 6 top graph), which is expected since it is anticipated to be proportional to the inverse of the RSSI. The maximum packet interval rate is measured to be $5.52$ Hz. The average RSSI and packet interval rates in TX location 1 is $1.51$ mW and $0.34$ Hz, respectively.

\begin{figure}[!ht]
   \centering
   \subfloat{\includegraphics[width=.4\textwidth]{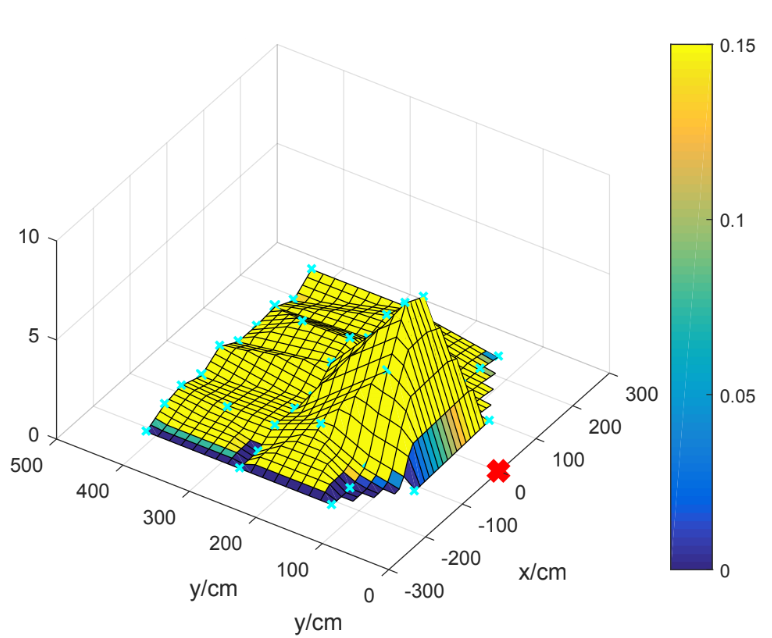}}
  \\
 \quad \subfloat{\includegraphics[width=.4\textwidth]{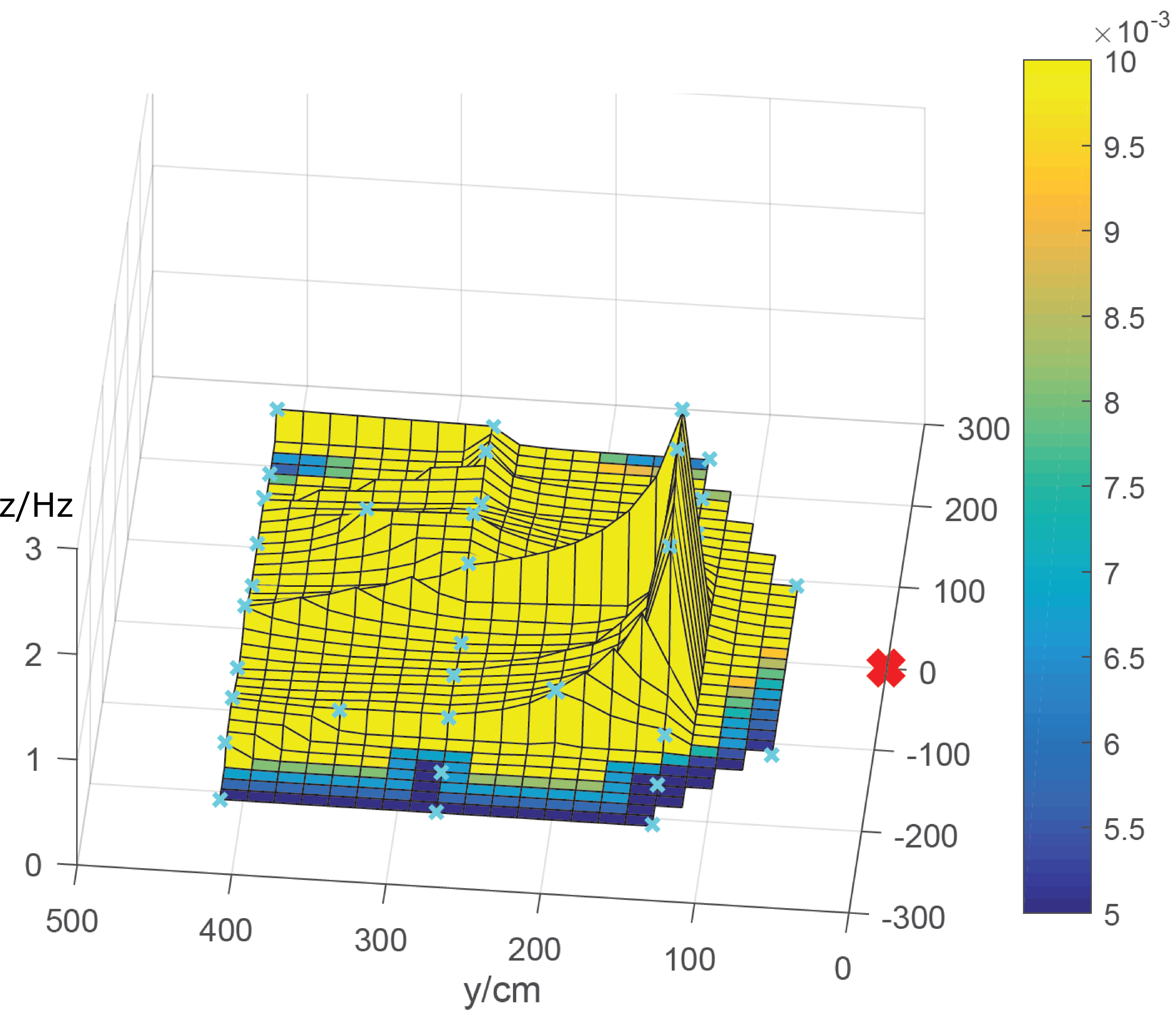}}
   \caption{Received power levels in mW (top) and packet rate frequency in Hz (bottom) for TX location 2. The red cross indicates the TX position. Note that the point of view is not the same for the plots. The colorbar is adjusted such that areas without reception are highlighted in blue. X and Y axis are interpolated by a factor of 4.}
   \label{TX2-results}
\end{figure}

As for TX location 2, the RSSI distribution is more widely dispersed around the cabin. The RSSI is significantly higher in the first row right in front of the TX compared with the other locations. Sufficient power to overcome the operational threshold for the low-power ZigBee TX is received in $28$ RX locations out of $36$. Although RX locations $31$ and $36$ are the closest locations to the power TX, the RSSI values at these locations are low due to beam misalignment. Similar to the TX location 1, the packet interval curve (Fig. 7 bottom graph) is steeper compared with the RSSI curve (Fig. 7 top graph). The maximum RSSI is measured to be $6.88$ mW and the maximum packet interval rate is $2.03$ Hz. The average RSSI and packet interval rates in TX location 2 is $0.98$ mW and $0.3$ Hz, respectively. 

The average RSSI rate is around $54$\% higher in TX location 1 compared with location 2. The graphs also show that the RSSI distribution is more evenly spread in TX location 2. Therefore, each TX configuration has certain advantages. TX location 1 can be preferred if high power consuming devices are targeted for wireless powering whereas the TX location 2 can be preferred to power more devices with a less number of TXs. 

As a conclusion, the results show that low-power sensors up to power consumption of $27.91$ mW along with a packet sending rate of $5.52$ Hz can be supported with the implemented WPT system onboard aircraft cabin. These results already imply the potential of WPT to wirelessly power certain use cases such as cabin monitoring sensors and low-power seat applications. 

\subsection{Multi-Tone Measurements in Laboratory Environment}
In these measurements, the RX board is fed with four different waveforms, all centered around $915$\,MHz with equal total average power. In every multi-tone signal, each sinusoidal component has the same amplitude and starting phase. The first multi-tone signal consists of two sinusoids with a spacing of $1$\,kHz and a Peak-to-Average Power Ratio (PAPR) of $3$\,dB. The second waveform uses three sinusoids with a spacing of $500$\,Hz and a PAPR of $4.8$ dB. The third multi-tone signal consists of five tones with a spacing of $250$\,Hz and a PAPR of $7$\,dB. All multi-tone signals have a bandwidth of $1$\,kHz but a variant number of sinusoidal components. 

\begin{figure}[htbp]
\centerline{\includegraphics[width=0.5\textwidth]{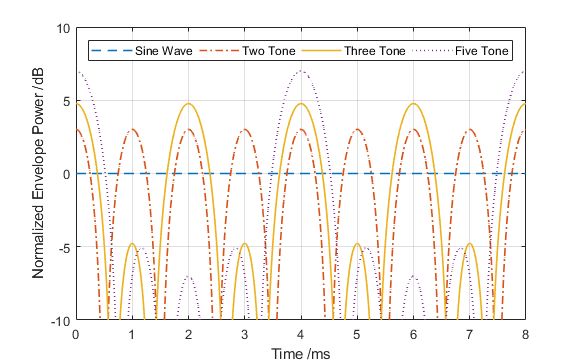}}
\caption{Normalized envelope power of the applied waveforms. The multi-tone waveforms are constructed with same amplitude and starting phase for all tones.}
\label{envelope}
\end{figure}

Figure \ref{envelope} shows the envelope of the three multi-tone waveforms and the single-tone waveform. Due to the non-linearity of the RF to DC conversion efficiency as reported in the Powercast manual\cite{b11}, it is expected that the average conversion efficiency may be improved for multi-tone beat signals.

\begin{figure}[htbp]
\centerline{\includegraphics[width=0.5\textwidth]{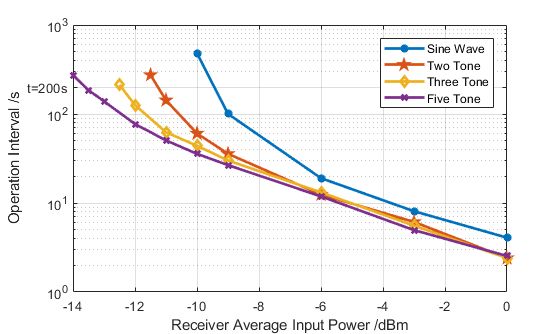}}
\caption{Operational interval as an inverse measure for the RF to DC conversion efficiency for different multi-tone signals with equal average power.}
\label{MT-results}
\end{figure}

The operational interval is recorded for each type of input signal for different input powers, starting at $0$\,dBm and decreasing to $-14$\,dBm. The results are shown in Fig. \ref{MT-results}. Most notably, the pure sine wave leads to the largest operational interval at all input powers. This discrepancy between the pure sine and the multi-tone signals rises for low input powers, where an improved RX sensitivity is most needed. The multi-tone signals perform similar above \mbox{-6\,dBm}. Below this value, the signals that contain more tones perform better. This leads to the RX board being able to operate down to lower average input powers. Comparing the sine wave with the five tone signal, a $4$\,dB smaller signal can still be received and used to power a sensor.  

\section{Conclusion and Discussion}

Overall, this study provides an insight regarding the feasibility of integrating RF-based WPT systems onboard the next-generation aircraft to wirelessly power the low-power sensors. 

As the usage of WPT inside an aircraft cabin requires careful regulatory considerations, it is significant to observe the unharmonized frequency as well as transmit power regulations across different nations. This can be a limiting factor to integrate WPT on cross-border flights. 

The measurements inside the mock-up cabin show the potential of a WPT system, which can provide power up to $27.91$ mW along with a $5.52$ Hz packet interval rate. These values already enable a number of cabin use cases to be powered up wirelessly with a WPT system. Additionally, laboratory measurements show potential ways to further enhance the RX sensitivity of the WPT system so that WPT coverage can be increased in the cabin. 

All in all, this study suggests that RF-based WPT systems can be feasible to implement in the next-generation aircraft to avoid power cable harnesses for certain low-power use cases. This will help further increase the overall aircraft efficiency. Future studies can elaborate on this topic to elaborate on the optimization of multiple WPT systems to cover an entire aircraft cabin. 

%\section{Appendix}
%All measured received power levels and packet frame rates can be found in the following Figure %\ref{appendix-table}.

\end{document}